\documentclass[prx,reprint,superscriptaddress]{revtex4-1}

\usepackage{graphicx}
\usepackage{soul}
\usepackage{amsmath}
\usepackage{xcolor}
\usepackage{braket}
\usepackage{siunitx}

\newcommand{\clocktransition}{$^1\mathrm{S}_0\,\rightarrow\,^3\mathrm{P}_1$ }

\newcommand{\qtty}[2]{#1\,\mathrm{ #2}}

\begin{document}

\title{Wavefront Curvature in Optical Atomic Beam Clocks}
\author{A. Strathearn}
\affiliation{ARC Centre for Engineered Quantum System, School of Mathematics and Physics, University of Queensland, Brisbane, QLD 4072, Australia}

\author{R. F. Offer}
\affiliation{Institute for Photonics and Advanced Sensing (IPAS) and School of Physical Sciences, University of Adelaide, Adelaide, South Australia 5005, Australia}

\author{A. P. Hilton}
\affiliation{Institute for Photonics and Advanced Sensing (IPAS) and School of Physical Sciences, University of Adelaide, Adelaide, South Australia 5005, Australia}

\author{E. Klantsataya}
\affiliation{Institute for Photonics and Advanced Sensing (IPAS) and School of Physical Sciences, University of Adelaide, Adelaide, South Australia 5005, Australia}

\author{A. N. Luiten}
\affiliation{Institute for Photonics and Advanced Sensing (IPAS) and School of Physical Sciences, University of Adelaide, Adelaide, South Australia 5005, Australia}

\author{R. P. Anderson}
\affiliation{School of Molecular Sciences, La Trobe University, PO box 199, Bendigo, Victoria
3552, Australia.}

\author{B. M. Sparkes}
\affiliation{Defence Science and Technology Group, Edinburgh, South Australia 5111, Australia}

\author{T. M. Stace}
\affiliation{ARC Centre for Engineered Quantum System, School of Mathematics and Physics, University of Queensland, Brisbane, QLD 4072, Australia}

\begin{abstract}
Atomic clocks provide a reproducible basis for our understanding of time and frequency. 
Recent demonstrations of compact optical clocks, employing thermal atomic beams, have achieved short-term fractional frequency instabilities of order $\num{e-16}$, competitive with the best international frequency standards available.
However, a serious challenge inherent in compact clocks is the necessarily smaller optical beams, which results in rapid variation in interrogating wavefronts. This can  cause inhomogeneous excitation of the thermal beam leading to long term drifts in the output frequency. 
Here we develop a model for Ramsey-Bord\'{e} interferometery using optical fields with curved wavefronts and simulate the $^{40}$Ca beam clock experiment described in [Olson \textit{et al.}, Phys. Rev. Lett. 123, 073202 (2019)]. The results of Olson \textit{et al.} have shown surprising and unexplained behaviour in the response of the atoms in the interrogation.
Our model predicts signals consistent with experimental data and can account for the significant sensitivity to laser geometry that was reported. 
We find the signal-to-noise ratio is maximised when the laser is uncollimated at the interrogation zones to minimise inhomogeneity, and also identify an optimal waist size determined by both laser inhomogeneity and the velocity distribution of the atomic beam. 
We investigate the shifts and stability of the clock frequency, showing that the Gouy phase is the primary source of frequency variations arising from laser geometry.

\end{abstract}

\maketitle

\section{Introduction}
State-of-the-art optical atomic clocks routinely achieve fractional frequency instabilities or uncertainties on the order of \num{e-18} \cite{Nicholson2015, Ushijima2015, Huntemann2016, McGrew2018, Brewer2019}, with recent work achieving uncertainties of order \num{e-21} \cite{Bothwell2022}. 
Such high precision experiments allow for ever improving tests of fundamental physics \cite{Godun2014, Huntemann2014,Bothwell2022}, more precise chronometric levelling-based geodesy \cite{Lisdat2016, McGrew2018} and the potential for the redefinition of the Standard International (SI) second \cite{Margolis2014, Riehle2015, Riehle2018}.
Optical atomic clocks take many forms and use a variety of interrogation protocols and techniques.
Single-ion clocks \cite{Huntemann2016, Brewer2019, Huang2020} and neutral atom lattice clocks \cite{Nicholson2015, Ushijima2015, Koller2017, Grotti2018, McGrew2018, Takamoto2020a,Bothwell2022} currently achieve the highest levels of long-term stability and accuracy.
These systems employ multiple stages of laser cooling and trapping to prepare a sample with zero net velocity and near-zero temperature, allowing for extremely long interaction times and ultra-high spectroscopic resolution.
Alternatively, thermal vapor cells provide very high number densities of the target atom \cite{Perrella2019,Newman2021} or molecule \cite{Zang2007,Schuldt2017,Doringshoff2019} with low system complexity.
The high signal to noise achievable in these devices provide excellent short-term stability in compact, robust packages.
Thermal atomic beam clocks \cite{McFerran2010,Marlow2021,PhysRevLett.123.073202} lie between these extremes, combining a relatively high atomic flux with kHz-scale spectroscopic resolution.
Already an industry standard for portable microwave clocks, these systems offer significantly reduced system complexity compared to trapped atom/ion clocks. This balance between size, weight and power (SWaP) and performance make thermal atomic beam clocks excellent candidates for high-performance portable devices \cite{Gutsch2019}.

Fundamentally, atomic clocks are devices that measure a particular atomic transition frequency. 
For simplicity we refer to the two states involved in such transitions as the ground and excited states. 
High-resolution measurement of microwave atomic clock transitions can be achieved via Ramsey spectroscopy \cite{PhysRev.78.695,PhysRevLett.90.150801,PhysRevLett.85.1622}, where two time-separated resonant microwave pulses are applied to the atomic sample. 
This is a form of atom interferometery \cite{Borde2002}, where the paths of the ground and excited states between the pulses interfere to produce oscillations in the measured population as the frequency of the applied field is varied. 
Viability of this interrogation protocol relies on the ground and excited state wavefunctions having spatial overlap at measurement, otherwise they cannot interfere.

At optical frequencies, absorption of photons is accompanied by significantly larger momentum transfer to the atom, approximately five orders of magnitude larger than for microwaves. 
This leads to increased displacement of the excited state wavefunction between pulses, hampering interference with the undisplaced ground state. 
To solve this problem, Ramsey-Bord\'{e} (RB) spectroscopy \cite{PhysRevA.30.1836} utilises a second pair of resonant pulses counter-propagating with respect to the first pair, correcting for wavefunction displacement and allowing interference to be observed. 
This pulse configuration also makes the interference fringe phase-insensitive to the first order Doppler effect (equivalently, spatial phase variation in the laser), making RB spectroscopy particularly suited for interrogation of atoms with significant velocity spread, for example in thermal atomic beams \cite{Morinaga1989,Riehle1991,McFerran2010,PhysRevLett.123.073202} or magneto-optical traps \cite{Wilpers2002,Wilpers2003, Degenhardt2005, Wilpers2006, Kisters1994}. An example transition used as the clock transition in these experiments is the \clocktransition transition of $^{40}$Ca atoms due to its narrow, but not too narrow, linewidth and insensitivity to external fields \cite{Morinaga1989}. 
On the theory side, improvements to bare two-pulse Ramsey spectroscopy have been made by using tailored pulses \cite{PhysRevA.82.011804,PhysRevLett.109.213002} and subsequent progress has been made in improving RB spectroscopy using quantum engineered composite pulses \cite{Zanon2022}.

The effect of wavefront curvature on laser-cooled neutral atom clocks, where the lasers are pulsed on and off into a cloud of stationary atoms, has been investigated previously \cite{Wilpers2003, Degenhardt2005, Wilpers2006, Friebe_2011}.
The thermal beam system differs significantly in that the interrogation zones are spatially separated, and therefore will exhibit differing levels of wave-front curvature determined by the exact layout of the apparatus.
In addition, atoms in a thermal beam traverse the lasers at high velocities and experience a rapidly varying degree of wavefront curvature within the interrogation zones themselves.

In this paper we investigate the effect that wavefront curvature has on a thermal beam clock, including on the signal strength and contrast, as well as frequency uncertainty.
We also test our model through comparison with the recent experimental work of Olson \textit{et al.} \cite{PhysRevLett.123.073202}, where it was speculated that wavefront curvature had significant impact on the observed signal.
These results will assist in determining the optimal beam waist sizes and locations for future compact thermal beam clocks, and allow calculation of the systematic uncertainties caused by these effects. 

In Sec. I we develop a theory of Ramsey-Bord\'{e} interferometry using a realistic Gaussian model for the laser. We highlight the difference between our theory and the standard theory using plane wave lasers and also compare the signal we calculate to the experimental signal reported in Olson \textit{et al.} \cite{PhysRevLett.123.073202}. In Sec. II we quantify the overall quality of the signal using Fisher information and use this to determine how the laser parameters can be optimised to give the best signal. In Sec. III we analyse sources of frequency shifts to the clock transition and quantify how instability in the laser geometry leads to instability in the frequency. In Sec. IV we conclude and discuss our results. 

\section{Ramsey-Bord\'{e} Interferometry}

\subsection{Ramsey-Bord\'{e} Signal}
\begin{figure}
\begin{centering}
\includegraphics[scale=0.6]{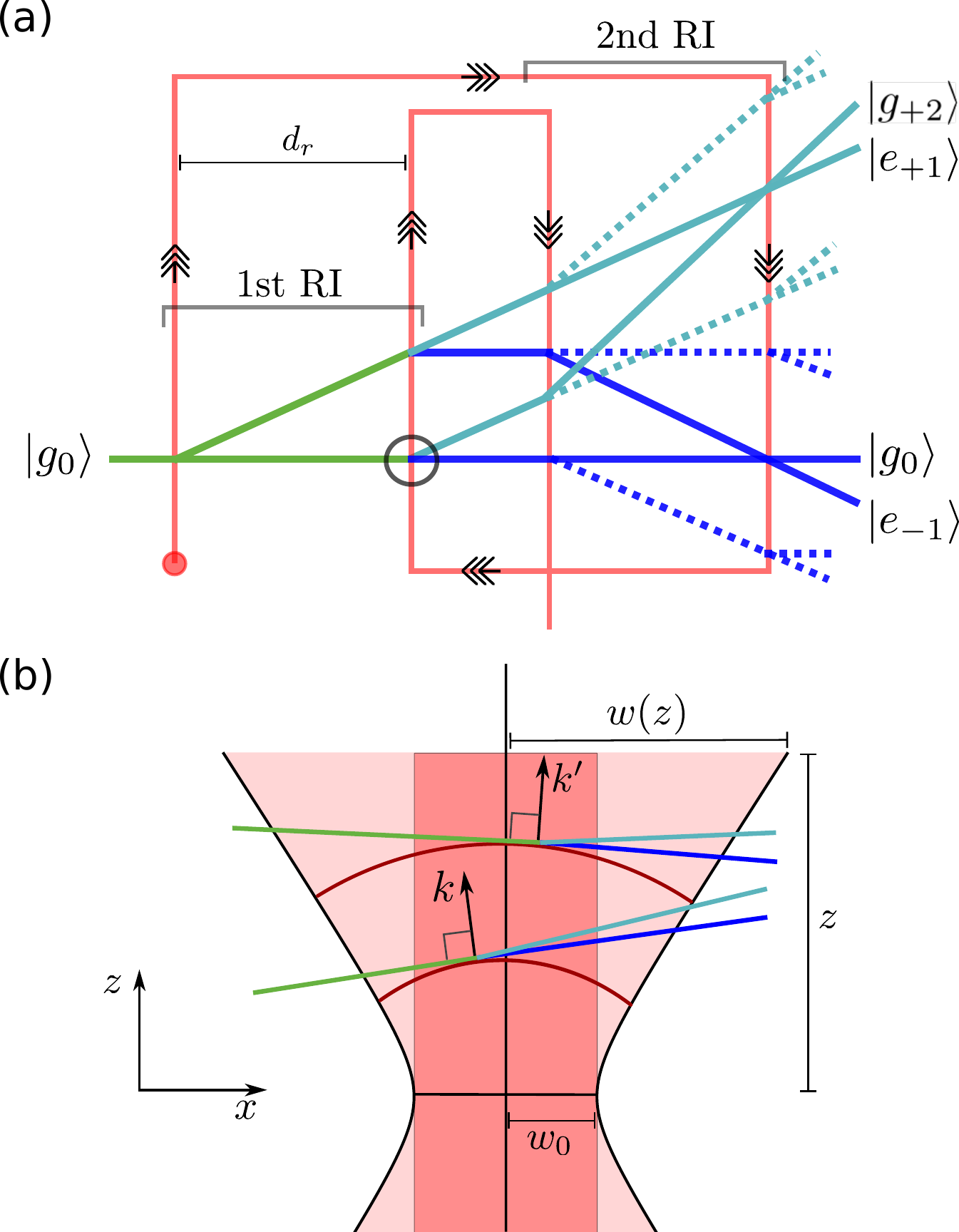}
\par\end{centering}
\caption{(a) Schematic of the Ramsey-Bord\'{e} interferometer.
The red line represents the single laser that has been folded to create four atom-laser interaction zones.
The atomic path starts in $\ket{g_0}$ and is split by the first interaction (green).
The paths split into upper (light blue) and lower (dark blue) recoil components, ending in $\ket{e_{+1}}$ and $\ket{e_{-1}}$.
Solid (dashed) lines indicate interfering (non-interfering) paths.
(b) Microscopic view of the interaction circled in (a) for two different atomic trajectories which are resonant with two different wavevectors, $k$ and $k'$. Red curves depict the curved wavefronts in the spreading laser field.
On resonance, atoms with are excited by the wavevectors normal to their trajectories that lie within a distance $w_0$ from the optical axis.
Wavevectors in the lightly shaded portion of the laser are far off resonance due to curvature and do not contribute.
}
\label{fig:1}
\end{figure}

An optical Ramsey interferometer (RI) could be realised by directing atoms through two copropagating lasers, each with wavevector $k$, separated by a distance $d_r$.
The phase of the lasers at each atom-laser interaction depends on the laser detuning, $\Delta$, from the atomic transition and the travel time, $T$, between lasers.
Defining the RI background, $a_+$, and envelope, $a_-$, functions, the excitation probability is
\begin{equation}\label{eq:Ram}
    p=a_++a_- \cos(\theta).
\end{equation}
The fringe phase, $\theta = \Delta T+\phi$, depends on the laser detuning, $\Delta$, from the atomic transition, the atomic travel time between lasers, $T$, and other contributions such as the laser Guoy phase which are included in $\phi$. As a function of $\Delta$, $p$ is a $1/T$ periodic oscillation with envelope $a_-$ superimposed upon an incoherent background $a_+$. 
For atoms with finite excited state lifetime $1/\gamma$, the background decomposes into contributions from atoms propagating in the ground and excited states between lasers, $a_+ \to a_g+e^{-\gamma T}a_e$, and since both contributions are required for interference the envelope decays, $a_- \to e^{-\gamma T} a_-$.
Atomic motion along the optical axis with velocity $v_z$ Doppler shifts the detuning $\Delta \to \Delta + v_zk$. 
Then if the spread in transverse displacements of atoms between the lasers due to the spread in $v_z$ is greater than a wavelength, $v_zk T>1$, the Ramsey fringes in Eq.~\eqref{eq:Ram} dephase and become unobservable. 
This is equivalent to saying that the fringes dephase if the spread in Doppler shifts $v_z k$ is greater than the fringe period.
Therefore for thermal atomic beams with spread in transverse displacements $\sim 1 \mathrm{mm}$ during the flight time, only microwave radiation can resolve fringes in a RI. 
Our description of Ramsey interferometery here neglects any spatial extent of atomic wavefunctions. 
A more fundamental problem with optical Ramsey interferometery is that momentum imparted by the lasers can lead to spatially non-overlapping wavefunctions of excited state atoms, such that no interference is observed even in the absence of the first order Doppler effect.

If one puts in place a second RI after the first, where the two lasers are counter-propagating with those in the first RI, the initial Ramsey fringes with Doppler shift $+v_zk$ form an envelope for the second Ramsey fringes with Doppler shift $-v_zk$, resulting in amplitude modulated fringes in the signal that contain a Doppler free oscillation mode. 
This is a Ramsey-Bord\'{e} (RB) interferometer, as depicted in Fig~\ref{fig:1}(a). 
Excitation by the laser is accompanied by momentum transfer to the atom of $\pm \hbar k$ along the wavevector of the laser, so the ground and excited states of the atom, $\ket{g_n}$ and $\ket{e_n}$, are indexed by the number, $n$, of imparted recoil momenta, $\hbar k$. 
For atoms initially in the state $\ket{g_0}$, the state space for the atoms throughout the RB interferometer is spanned by the basis $\{\ket{g_0},\ket{e_1},\ket{e_{-1}}, \ket{g_2}$\}. 
In Fig~\ref{fig:1}(a) we show the possible paths the atoms can take through the interferometer to transition from the initial ground state $\ket{g_0}$ to the upper, $\{\ket{e_{1}},\ket{g_2}\}$, and lower, $\{\ket{g_0},\ket{e_{-1}}\}$, recoil states.
The $n$th recoil states have additional kinetic energy $n^2\delta = n^2\hbar k^2/2m$, so the detuning for transitions $\ket{g_n} \to \ket{e_{n\pm1}}$ is shifted by $(1\pm 2n) \delta$. 
In addition to being Doppler free, the RB interferometer is also insensitive to the spatial displacement of atomic wavefunctions due to recoil. 
This can be seen in Fig~\ref{fig:1}(a), where interference occurs between the pairs of atomic paths that end at the same point at the final laser. 

In the following, we denote the signal parameters relating to the second RI in the RB with a prime e.g. the excitation probability is $p'$. 
The lower recoil signal is $p_L =(1-p)p'$ since atoms transition $\ket{g_0} \to \ket{g_0}$ through the first RI with probability $(1-p)$ then transitions $\ket{g_0} \to \ket{e_{-1}}$ with probability $p'$ through the second. 
The upper recoil signal is $p_U =p(1-p')$ since atoms first transition $\ket{g_0} \to \ket{e_1}$ with probability $p$ then transition $\ket{e_1} \to \ket{e_{1}}$ with probability $1-p'$. 
In both $p_L$ and $p_U$ are fringes from each RI with Doppler shifted phases, $\theta$ and $\theta'$, as well as oscillations $\propto \cos(\theta)\cos(\theta')$, that decompose into a mode with Doppler shifted phase $\theta-\theta'$, and a Doppler free mode with phase $\theta+\theta'$. 
With reference to Eq.~\eqref{eq:Ram}, and retaining only the Doppler free fringes, the signal components of the RB are
\begin{align}\label{eq:RBL}
    p_L&=a'_+(1-a_+ )-\frac{1}{2}a_-a'_- \cos(\theta_L) \\ \label{eq:RBU}
    p_U&=a_+(1-a'_+ )-\frac{1}{2}a_-a'_- \cos(\theta_U).
\end{align}
We stress that the second RI signal parameters $a_\pm'$ and $\theta'$ are functions of different detunings for $p_L$ and $p_U$. 
Importantly, we have $\theta_{L/R}=\theta+\theta'=2T(\Delta\pm\delta)+\phi+\phi'$ which exhibits either a lower (L), $-\delta$, or upper (R), $+\delta$, recoil shift. 
If the Ramsey times, $T$ and $T'$, between each pair of lasers are not equal then there is a residual first order Doppler shift, $(T-T')v_zk$, in the fringes.  
Due to larger momentum states involved in the upper recoil transitions the atom has net axial motion even when ${v_z=0}$. 
This results in a net first order Doppler shift $2\delta$ to the laser detunings in each RI around the upper recoil shift, $\Delta+\delta \pm(2\delta +v_z k)$. 
This constant Doppler shift cancels from the fringe phase by design, but means that when tuned to the upper recoil resonance, $\Delta=-\delta$, the lasers are not actually resonant with the transition during either RI so there is a loss in pulse area and resulting decrease in excitation probability.  
This implies there is an intrinsic difference in the quality of the upper and lower recoil fringes.

\subsection{Gaussian Laser Model}
To evaluate Eqs.~\eqref{eq:RBL} and \eqref{eq:RBU} we require expressions for the background and fringe envelope functions, $a_\pm$ and $a_\pm'$, and the overall phase shift, $\phi$. We define the pulse area as the polar angle of rotation of the Bloch vector of the internal atomic degrees of freedom that is induced by the interaction with laser. Writing the pulse area imparted on the atom by the $i$th laser as $2\Phi_i$, the backgrounds and fringe envelopes of the RIs are defined as
\begin{align}\label{eq:apm}
    a_\pm &=\frac{1}{2}\left(\sin^2(\Phi_1+\Phi_2) \pm \sin^2(\Phi_1-\Phi_2)\right) \\
    \label{eq:adpm}
    a'_\pm &=\frac{1}{2}\left(\sin^2(\Phi_3+\Phi_4) \pm \sin^2(\Phi_3-\Phi_4)\right)
\end{align}
In this form it is clear that the Ramsey fringe contrast $a_-/a_+$ is maximised for homogeneous pulse areas, i.e.  ${\Phi_1=\Phi_2}$. 
The phase shift, $\phi+\phi'=\psi_1-\psi_2+\psi_3-\psi_4$, in the RB fringes is determined by the laser phases, $\psi_i$. To find $\Phi_i$ and $\psi_i$ we model the laser as a propagating transverse Gaussian mode with wavevector $k$ and waist radius $w$, such that the Rayleigh range is $z_R=kw^2/2$. 
Using the rotating wave approximation, valid at optical frequencies, the Hamiltonian that governs the dynamics of each atom-laser interaction in the rotating frame is $\tilde{H}(t)=E(t)e^{-i \Delta t}\ket{e_{n\pm1}}\bra{g_n} +h.c.$, 
where $\Delta$ is the detuning and the laser field amplitude is $E(t)=\Omega(t)e^{i\phi(t)}$ with
\begin{align}
\Omega(t)&= \frac{v_m A}{2 \sqrt{\pi} w(z)}\exp\left( -\frac{v^2 t^2}{w(z)^2} \right)\\ 
\phi(t) & =kz+k\frac{v^2t^2}{2R}-\mathrm{atan}(z/z_R).
    \label{eq:phase}
\end{align}
Here, $w(z)=w_0\sqrt{1+z^2/z_R^2}$ and $R=z+z_R^2/z$ are the physical width and wavefront radius of curvature a distance $z$ away from the waist along the laser axis. 
The amplitude $\Omega(t)$ is normalised such that atoms intersecting the waist with velocity $v=v_m$ perpendicular to the optical axis receives the target pulse area $2\Phi =A$. 
Away from the waist and for different velocities the actual pulse area differs from the target $2\Phi \ne A$.
We have neglected axial atomic motion over the transit time $v_z\tau/z_R \ll 1$. 
In principle, the curved wavefront of the laser field resulting from the second term in Eq.~\eqref{eq:phase} gives a continuum of possible atomic recoil momenta. 
We only model recoil along the optical axis, and assuming the spatial atomic wavefunction is localised well within a Rayleigh range we can neglect corrections to the bare recoil momentum $\pm \hbar k$.

The propagator for atomic evolution through the $j$th laser from time $t=-\tau$ to $t=\tau$ factorises, $U_j=U_\tau \tilde{U}_j U_\tau$, where $U_\tau$ is the free atomic propagator over time $\tau$ and $\tilde{U}_j=\mathcal{T}\exp(-i\int_{-\tau}^\tau \tilde{H}_j(t)dt)$ is the interaction picture propagator, with $\tilde{H}_j(t)$ as defined above.
In this form, we can think of the atom as propagating freely everywhere except at the optical axis, $t=0$, where it experiences an instantaneous interaction that generates $\tilde{U}$. 
Then the excitation probability through the first RI, Eq.~\eqref{eq:Ram}, is  $p= | \bra{e_1}\tilde{U}_2 U_T \tilde{U}_1 \ket{g_0}|^2$, where $\tilde{U}_j$ describes the $j$th atom-laser interaction. 
The operator $\tilde{U}_j$ can be calculated numerically by sequentially multiplying $n$ short time propagators $\exp(-i\int_{t_k}^{t_{k+1}} \tilde{H}(t)dt)$, with $1\leq k < n$. This is a form of Trotter splitting, and since $\tilde{H}_j(t)$ is Gaussian the integral in each propagator can be evaluated in terms of error functions. 
We find a simple approximate form for $\tilde{U}$ by taking $n=1$, recovering the first order Magnus expansion \cite{Magnus1954} which is $\log P \approx -i\int_{-\tau}^\tau \tilde{H}(t)dt$. 
The finite width of the field, $E(t)$, means we can take $\tau \to \infty $ to obtain $\tilde{U}$ in terms of $\tilde{E}(\Delta)=\Phi e^{-i \psi}$, the Fourier transform of $E(t)$, where $2\Phi$ and $\psi$ are the effective pulse area and laser phase. Evaluating $\tilde{E}(\Delta)$, we find
\begin{align}
    \Phi &= \frac{A}{2}\frac{v_m}{v}\sqrt{\frac{w_0}{w(z)}} \exp\left(-\frac{1}{4} \Delta^2 \tau_0^2\right) \label{eq:MEPA} \\
    \psi &= kz\left(1- \frac{1}{2}\frac{\Delta^2}{k^2v^2}\right)+\frac{1}{2}\arctan\left(\frac{z}{z_r}\right) \label{eq:MEPh},
\end{align}
which we can use to evaluate Eqs.~\eqref{eq:apm} and \eqref{eq:adpm}.
Taking the Fourier transform of the optical field $E(t)$ to obtain the expressions above has identifies the resonant wavevectors in the laser when it is detuned by $\Delta$. 
These wavevectors lie in a constant region with radius the size of the waist, $w_0$, along the full optical axis, shown as a dark shaded region in Fig.~\ref{fig:1}(b), even away from the waist where the physical laser radius is $w(z)>w_0$.
This means the effective atomic transit time, $\tau_0=w_0/v$, is constant along the optical axis and $\Phi$ has Gaussian transit broadening $\propto 1/\tau_0$. 

As the laser field diverges from the waist the intensity in the resonant paraxial region described above decreases. 
The resulting loss in pulse area is reflected in Eq.~\eqref{eq:MEPA} through the factor $\sqrt{w_0/w(z)}=(1+(z/z_r))^{-1/4}$. 
The resonant wavevector in a detuned laser is rotated off-axis by a small angle $\sim \mp \Delta/kv$ since the component along the atomic trajectory is Doppler shifted by $kv\sin(\Delta/kv) \approx \Delta$, compensating for the detuning. 
The $z$-component of the resonant wavevector is $k_z \approx k\cos(\Delta/kv)$, and the first term in Eq.~\eqref{eq:MEPh} is the small angle approximation of the spatial phase $k_z z$. For atoms with axial Doppler shift $v_zk$ and small detuning the resonant wavevector is rotated by $\sim v_z/v$, making it normal to the atomic trajectory.
We sketch this in Fig.~\ref{fig:1}(b). 
Since the atom is being excited a distance $\sim z\sin(v_z/v)$ off axis we find there is a correction to the free propagation time of the atom given by $z v_z/v^2$. 
The $\arctan$ term is the Gouy phase in Eq.~\eqref{eq:MEPh}.

\subsection{Comparison With A Plane Wave Model }
To calculate the RB signal in a realistic setting we assume experimental conditions similar to those described in \cite{PhysRevLett.123.073202}, where they address the $\qtty{657}{nm}$ $^1\mathrm{S}_0\,\rightarrow\,^3\mathrm{P}_1$ transition of $^{40}$Ca atoms. The recoil shift in this case is $\delta \approx \qtty{11.5}{kHz}$.
Taking the origin of the optical path as the first atom-laser interaction zone, the positions of interaction zones along the optical axis are $\{l_1,l_2,l_3,l_4\}=\{\qtty{0}{cm},\qtty{51}{cm},\qtty{77}{cm},\qtty{30}{cm} \}$. We note that the folded geometry of the laser, as shown in Fig.~\ref{fig:1}(a), means the distance the atom travels to the interaction zones does not linearly increase with distance along the optical axis.   
The position and size of the waist are variables $l_w$ and $w_0$, and the distance of the atom from the waist at the $i$th interaction is $z_i=l_i-l_w$. 
The spatial separation of the atom-laser interactions in each RI is $d_r=\qtty{9}{cm}$, and we neglect the separation between second and third laser. 
The $^{40}$Ca clock transition has wavelength $\qtty{657}{nm}$ so waist radii in the range $\qtty{0.125}{mm}<w< \qtty{0.3}{mm}$ give Rayleigh ranges $\qtty{7.5}{cm}<z_R<\qtty{43}{cm}$.

To find the velocity averaged signal that is observed in experiment we assume the atomic beam has a thermal longitudinal velocity distribution, ${\rho_v \propto v^3 \exp(-m v^2/k_B T)}$, with temperature $T=\qtty{625}{K}$ giving a mean velocity $v_m\approx\qtty{610}{m/s}$ for $^{40}$Ca.
The transverse velocity distribution is determined by apertures that shape the atomic beam. We assume a Gaussian transverse velocity distribution, $\rho_{v_z} \propto \exp(-(v_z-v_0)^2/v_w^2)$, with width $v_w \propto v d_a$ for an atomic beam shaped by apertures of diameter $d_a$. 
The mean transverse velocity $\langle v_z \rangle = v_0 = v \sin(\alpha)$ accounts for angular misalignment $\alpha$ between atomic and laser beams.

To calculate the velocity averaged signal we compute the signal for various $(v,v_z)$, and integrate over the distributions. Typically we find convergence of the final result using $\sim 400$ equally spaced values of $v$ between $v\approx \qtty{100}{m/s}$ and $v \approx \qtty{2200}{m/s}$ and $\sim 40$ values of $v_z$ with $|v_z| \leq 3v_w$. We denote velocity averaged quantities with angular brackets.

The signal measured in \cite{PhysRevLett.123.073202} is the combined upper and lower recoil signals, $p_L+p_U$, averaged over the velocity distributions of the atomic beam.
We parameterise the average signal as
\begin{equation}\label{eq:P}
    P = \langle p_L \rangle +\langle p_U \rangle = b(1+c),
\end{equation}
where
\begin{equation}
b = \langle a_+' \rangle_L + \langle a_+ \rangle_U -\langle a_+a_+' \rangle_L -\langle a_+a_+' \rangle_U
\end{equation}
is the total background and
\begin{equation}
c = \frac{-1}{2b}\left(\langle a_- a_-' \cos(\theta_L) \rangle_L +\langle a_- a_-' \cos(\theta_U) \rangle_U \right)
\end{equation}
is the contrast function. We have used subscript $L/U$ to distinguish contributions from the upper and lower recoil atoms which have different recoil shifts as a function of detuning.
The first two terms in $b$ are the individual background contributions from each RI within the RB. The contrast function $c$ is a sum of the averaged upper and lower recoil fringes with amplitude given as a fraction of $b$. In Appendix \ref{app:RI} we show the functional forms of $a_\pm$ for atoms different velocities in the case of homogeneous pulse areas where $a_+=a_- \equiv a$.

\begin{figure}
\begin{centering}
\includegraphics[scale=1]{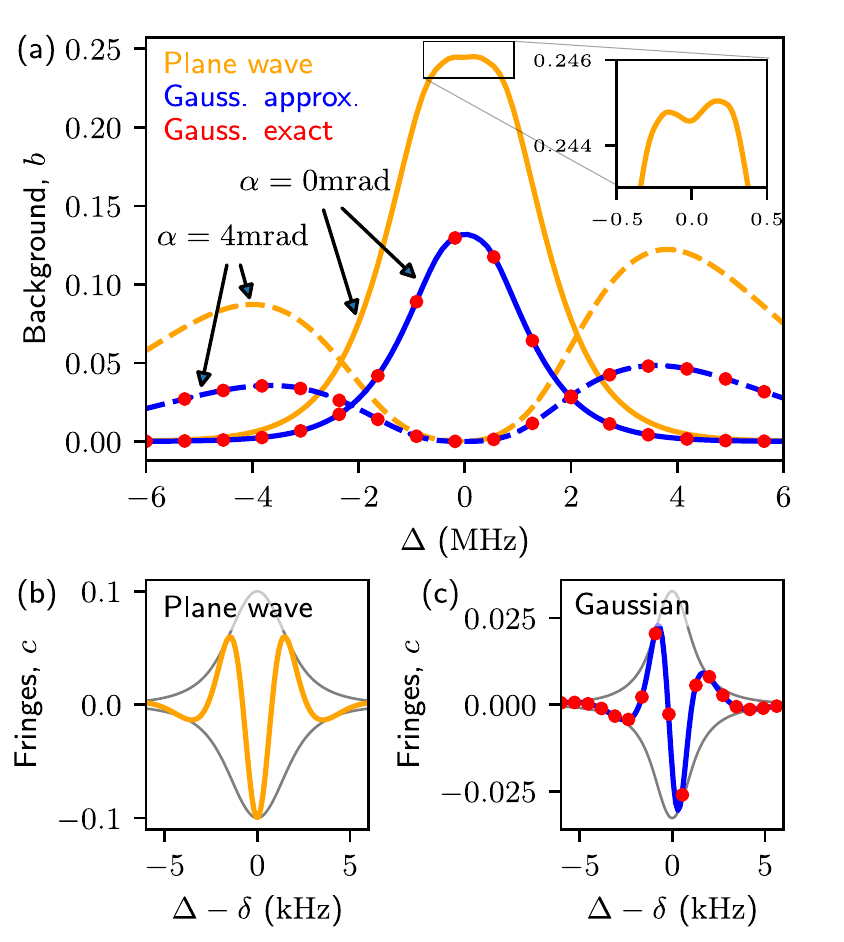}
\par\end{centering}
\caption{\label{fig:2} RB signal using plane wave and Gaussian lasers. Red points indicate results of exact numerical calculations. 
(a) Background signal for atomic beam perfectly perpendicular to the lasers, $\alpha=0$, and tilted, $\alpha=\qtty{4}{mrad}$. 
We highlight two different sources of asymmetry: misalignment of the symmetric Doppler background due to recoil, shown in the inset, and the splitting of the single peak into decaying and decay-free parts when the atomic beam is tilted. 
(b-c) Lower recoil fringe contrast function when the atomic beam is aligned for plane wave, (b), and Gaussian, (c), lasers. We have shifted the detuning by the recoil frequency $\delta = \qtty{11.6}{kHz}$ and solid black lines are the fringe envelopes.}
\end{figure}

In Fig.~\ref{fig:2}, we plot $b$ and $c$ using a Gaussian laser with $z_R=\qtty{7.5}{cm}$, and with a plane wave laser. The plane wave laser model is found simply by neglecting the second and third terms in Eq.~\eqref{eq:phase}. We have taken the target pulse area to be $A=\pi/3$ which for the Gaussian lasers decays along the optical axis according to Eq~\eqref{eq:MEPA}.
For the plane wave laser the target pulse area is achieved all along the optical axis.
For these simulations we have used $v_w = v/v_m \times \qtty{0.5}{m/s} $, consistent with a narrow atomic beam $d_a \sim \qtty{0.5}{mm}$ and giving Doppler broadening similar to the laser transit time broadening $\sim \qtty{1}{MHz}$. 
All curves in Fig.~\ref{fig:2} are computed using the analytic results Eqs. \eqref{eq:MEPA} and \eqref{eq:MEPh}, while numerically exact results are shown by red points. 
We found convergence by splitting the exact time dependent atom-laser propagator into $n=10$ Trotter timesteps, providing very little correction to the $n=1$ analytic results which are thus essentially exact in this regime. 

Due to the intrinsic Doppler shift of the upper recoil atoms, both upper and lower recoil Ramsey backgrounds, $a_+$ and $a_+'$, align with the lower recoil resonance, $\Delta=\delta=\qtty{11.6}{kHz}$, and are Doppler shifted in opposite directions around this point. 
In Fig.~\ref{fig:2}(a) we plot the background, $b$ for when the atomic beam is perfectly perpendicular to the lasers with no net Doppler shift, $\alpha=0$, and for when the beam is tilted by $\alpha = \qtty{4}{mrad}$ giving a net Doppler shift $\sim \qtty{4}{MHz}$. 
The backgrounds of the individual RIs, $\langle a_+ \rangle$ and $\langle a_+' \rangle $, are the dominant contributions to the total RB background $b$.
When the atomic beam is aligned perpendicular to the lasers, $\langle a_+ \rangle$ and $\langle a_+' \rangle $ coincide at the lower recoil frequency, giving a single peak at $\Delta=\delta=\qtty{11.6}{kHz}$. When the atomic beam is tilted, $\langle a_+ \rangle$ and $\langle a_+' \rangle $ are Doppler shifted in opposite directions by $\sim \qtty{4}{MHz}$, making them individually resolvable as peaks at $\Delta = \pm \qtty{4}{MHz}$.
Atoms contributing to the upper recoil Ramsey background $\langle a_+ \rangle$ are excited in the first RI and decay for a time $T$ while passing through the second RI before detection, giving rise to the smaller of the peaks at $\sim -\qtty{4}{MHz}$. 
Atoms contributing to the lower recoil Ramsey background $\langle a_+' \rangle$ remain in the ground state through the first RI and are detected immediately after the second, giving a larger decay free peak at $\sim +\qtty{4}{MHz}$. 
This decay induced asymmetry between Ramsey backgrounds implies that even small angular misalignment can lead to asymmetry in the total background $b$.

In addition to the Doppler sensitive terms $\langle a_+ \rangle$ and $\langle a_+'\rangle$ in the background there are two smaller terms each of the form $-\langle a_+a_+' \rangle $ that are Doppler shift free, centered at the upper and lower recoil resonances $\Delta = \pm \delta$. 
When the atomic beam is perpendicular to the lasers the Doppler contributions $\langle a_+ \rangle$ and $a_+'$ coincide to give a single peak and the $-\langle a_+a_+' \rangle $ terms manifest as a Lamb dip in this peak, as shown in the inset of Fig.~\ref{fig:2}(a). This dip is not visible for the Gaussian laser due to the loss in pulse area away from the waist, but would become visible by increasing the laser intensity.
Since $\langle a_+ \rangle$ and $\langle a_+'\rangle $ are symmetric around $\Delta=\delta$ but the $-\langle a_+a_+' \rangle $ terms are symmetric around $\Delta=0$, the overall dip is shifted from the center of the background by $- \delta$ leading to asymmetry. 
We stress that this is different from the decay induced asymmetry that becomes visible when tilting the atomic beam: the asymmetry shown in the inset of Fig \ref{fig:2}(a) is intrinsic to the design of the RB interferometer.

The signal has fringes that sit atop the background in Fig.~\ref{fig:2}(a), which we plot for plane wave and Gaussian lasers in Fig.~\ref{fig:2}(b-c), assuming the atomic beam is perpendicular to the lasers, $\alpha=0$. The fringe envelopes are found by taking the absolute value of $-\langle a_- a_-' e^{i \theta_{U/L}}\rangle/(2b)$, and are marked with black lines.
The envelopes are approximately Gaussian, with width inversely proportional to the width of the of the velocity distribution of contributing atoms. 
Compared to the plane wave case, Fig.~\ref{fig:2}(b),  the envelope and fringes are narrower when using a Gaussian laser, Fig.~\ref{fig:2}(c), because the reduced pulse areas bias the probability that an atom contributes towards slower atoms in the Boltzmann distribution where it is more flat. 
Both the loss in pulse area and inhomoegeneity of the pulse areas when using the Gaussian laser lead to a lower contrast than if a plane wave was used.

\begin{figure}
\begin{centering}
\includegraphics[scale=1]{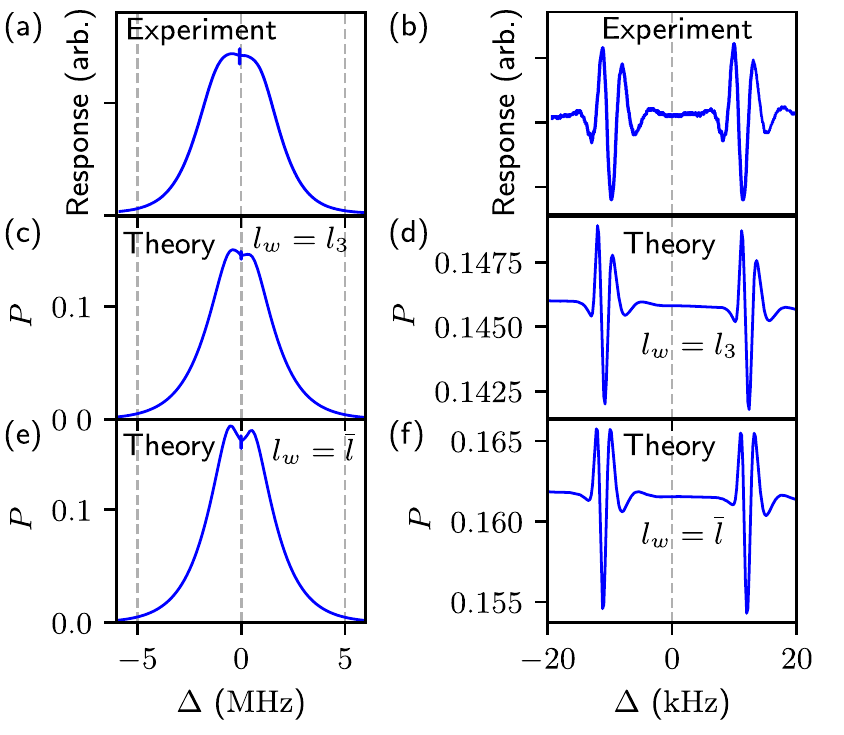}
\par\end{centering}
\caption{\label{fig:3} Experimental results from \citet{PhysRevLett.123.073202}, showing (a) the broad background and (b) the central fringes.
Numerical calculations showing the signal $P$ for different waist positions for: (c) the waist positioned at the third atom-laser interaction, $l_w=l_3$, showing the broad background
(d) and the fringes for the same waist position, and for; (e) the waist positioned at the average interaction position, $l_w=\overline{l}$, showing the broad background and (f) the fringes for the same waist position.
We use $w=\qtty{0.125}{mm}$ and target pulse area $A=\pi/2.5$. }
\end{figure} 

Here we have neglected the spatial laser phases, $kz_i$, at each interaction so the plane wave fringes in Fig.~\ref{fig:2}(b) experience no phase shift, yielding a central minimum fringe symmetric around resonance.
The combination of spatial phase fluctuations, atomic time of flight corrections, and Gouy phase in the  Gaussian laser has lead to a frequency shift $\sim \qtty{0.5}{kHz}$ of the central fringe in  Fig.~\ref{fig:2}(c) away from the center of the envelope, making the fringe profile asymmetric.
Velocity dependent phase fluctuation caused by wavefront curvature slightly shift the Gaussian envelope itself but this effect is too small to be seen in Fig.~\ref{fig:2}(c). 
We further discuss the phase shifts arising from using a Gaussian laser later.

\subsection{Comparison With Experiment}
Additional datasets \cite{OlsonThesis} from the RB experiment described in \cite{PhysRevLett.123.073202} show Doppler backgrounds whose asymmetry and dip visibility show significant variation. 
Such variation can be accounted for by the effects of wavefront curvature and angular misalignment of the atomic beam described above.
In Fig.~\ref{fig:3} we compare the data published in \cite{PhysRevLett.123.073202}(a-b) to simulations using our model (c-f). We use a waist size $w_0=\qtty{0.13}{mm}$ and two different waist positions: $l_w=l_3$, such that the waist is aligned with the third atom-laser interaction, and $l_w=\overline{l}$, where $\overline{l}=\sum_i l_i/4=\qtty{39.5}{cm}$ is the mean position of the interaction zones along the optical axis. 
Here we plot the full velocity averaged signal $P$ as would be measured rather than its separate components. In these simulations we also account for the relativistic Doppler effect by letting $\Delta \to \gamma(\omega_c +\Delta)-\omega_c$, where $\omega_c$ is the bare clock transition frequency and $\gamma = 1/\sqrt{1-v^2/c^2}$, with $c$ the speed of light. In this section we use a cubic Boltzmann distribution for longitudinal velocities, $\rho_v \propto v^3$, to get a better fit with experimental data but throughout the rest of the paper we use the standard Boltzmann distribution $\rho_v \propto v^2$.

The asymmetry of the central Doppler peak in the experimental data Fig.~\,\ref{fig:3}(a) is weighted in the wrong direction for it to be attributable to the intrinsic recoil asymmetry, so we assume a tilt of the atomic beam $\alpha = -160\, \mu \mathrm{rad}$ which produces a similar degree of background asymmetry in our data. 
This tilt of the atomic beams splits the background peaks by $\sim \qtty{0.3}{MHz}$. 
Our broad transverse velocity distribution $v_w = v/v_m \times \qtty{1.2}{m/s} $, consistent with an atomic beam of width $ d_a\sim \qtty{1}{mm}$, gives Doppler broadening $\sim\qtty{2}{MHz}$ so the two background peaks are not separately resolvable but do result in asymmetry, evident in Figs.\,\ref{fig:3}(c) and(e).

The Lamb dips in our numerical results are mostly due to the Doppler free contributions to the background, $-\langle a_+a_+' \rangle $, but are also partly due to the slower atoms receiving pulse areas $2 \Phi>\pi$, which results in dips in the bare Ramsey backgrounds $\langle a_+ \rangle$ and $\langle a_+'\rangle $. 
The target pulse area for average velocity atoms was chosen as $A=\pi/2.5$ instead of the ideal $A=\pi/2$ to reduce the Lamb dip to more closely match the experimental data, where the overall dip is extremely small.
The dip is more prominent when the waist is positioned at the symmetric point $l_w=\overline{l}$ since this positions maximises the average pulse area over the four interaction zones.
We found that lowering the target pulse area, $A<\pi/2.5$, can remove the dip to give an even better fit to the experimental background signal (not shown here) but this resulted in a contrast at the central fringes of $c < 0.01$, much less than the experimental contrast $c \sim 0.09$ reported in \cite{PhysRevLett.123.073202}.
In the high laser intensity regime, many atoms begin to experience pulse areas $2\Phi>\pi$, the Lamb dip becomes very large, and we expect our perturbative analytic solution, Eqs.~\eqref{eq:MEPA} and \eqref{eq:MEPh}, to break down.

The central fringes we calculate in Fig.~\,\ref{fig:3}(d) and (f) have contrasts $c \sim 0.02$ and $c \sim 0.03$, respectively, which falls short of the contrast $c \sim 0.09$ reported in \cite{PhysRevLett.123.073202}. As mentioned, our contrasts can be increased to the experimental value by increasing the laser intensity, and thus pulse area, but at the expense of producing a Lamb dip much larger than that observed. The narrowness of our fringes shown in Figs.\,\ref{fig:3}(d) and (f) compared to the experimental fringes in Fig.~\,\ref{fig:3}(b) is also due to our choice of small pulse area, such that the greatest contributions to the fringes come from atoms with slower than average velocity. For comparison, the target pulse area used in\cite{PhysRevLett.123.073202} is nominally the ideal value $A=\pi/2$ \cite{OlsonThesis}.

We tune the degree of fringe asymmetry in Figs.\,\ref{fig:3}(d) and (f) by shifting the optical position of the first atom-laser interaction by a fraction of the laser wavelength, $l_1 \rightarrow l_1 - 0.65 \lambda$.
This induces a $0.65 \times 2\pi$ phase shift of the fringes which have also been shifted away from the centre of the fringe envelope due to the relativistic Doppler effect and wavefront curvature.
In Fig.~\,\ref{fig:3}(d) this gives a similar degree of asymmetry as seen in the experimental data, Fig.~\,\ref{fig:3}(b), while in Figs.\,\ref{fig:3}(f) the different positions of the laser waist leads to more symmetric fringes.
The difference in fringe profile between Figs.\,\ref{fig:3}(d) and (f) is due to a combination of the different Guoy and wavefront phase shifts in the two cases, but also due to different average pulse areas.
Difference in pulse area leads to atoms with different velocities contributing more to the signal and if faster atoms contribute more then the relativistic Doppler shift is more prominent.
The relativistic Doppler shift leads to both phase shifts of the fringed and asymmetry in the fringe envelope itself, which is evident in Fig.~\,\ref{fig:3}(f).
These effects are complicated to describe accurately so for the remainder of this paper neglect the relativistic Doppler shift and analyse the effects of wavefront curvature in isolation.

\section{Brightness, Contrast, and Fisher Information}

Ideally, the RB signal, Eq.~\eqref{eq:P}, should arise from a large total number of measured atoms $N$, giving a large background contribution at the central fringe, which we call the brightness $b_0$, with a large fraction $N_f/N \propto b_0c_0 $ of these atoms producing fringes that are used to measure the clock laser detuning, $\Delta$. We define the fringe contrast, $c_0$, as the maximum possible contrast of the central fringe, i.e. the maximum amplitude of the fringe envelope, as a fraction of the brightness.

We assume atom shot noise is the primary source of measurement noise, in which case an estimate of the signal-to-noise ratio is $\sim N_f/\sqrt{N} \sim \sqrt{b_0}c_0$. 
More formally, the signal quality can be quantified by its Fisher information, $F$, which characterises the variance in the measured value of $\Delta$ as having a lower bound $\sigma_{\Delta} \geq 1/F$, the Cramer-Rao bound \cite{Rao1992}. 
The standard deviation in the measured detuning is $ \sim 1/\sqrt{F}$, and the signal-to-noise ratio is therefore $\propto \sqrt{F}$ which is maximised when $F$ is maximised. 
Here we analyse only the lower recoil fringes since, other than the small intrinsic degradation of contrast due to the upper recoil atoms having $\pm2\delta$ Doppler shift, both sets of fringes are identical. Concretely, this means $b_0=b|_{\Delta=\delta}$ and $c_0 = c|_{\Delta=\delta}$.  

\begin{figure}
\begin{centering}
\includegraphics[scale=1]{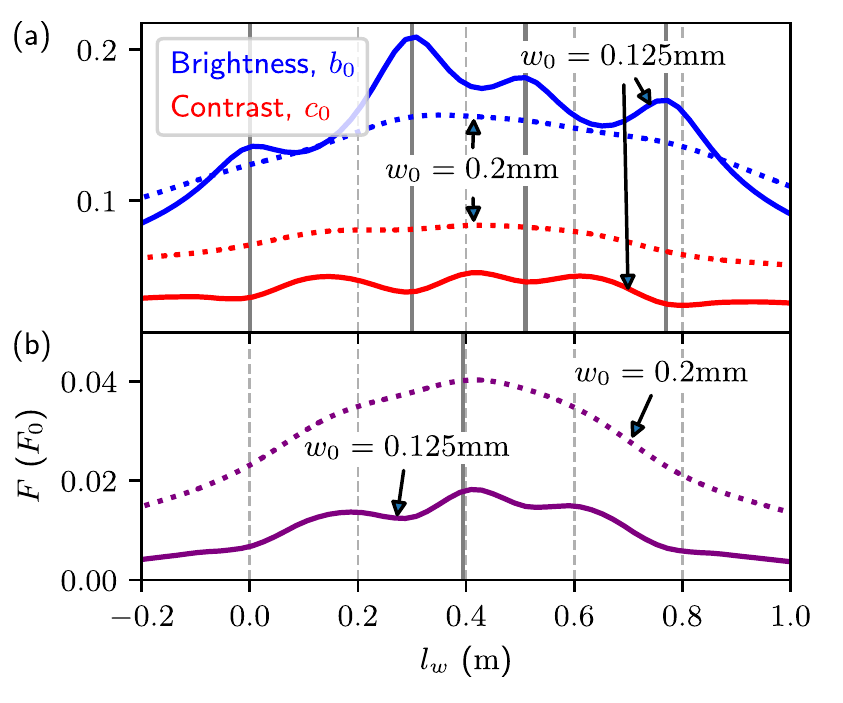}
\par\end{centering}
\caption{\label{fig:4} (a) The brightness, $b_0$, and contrast, $c_0$, at the central lower recoil fringe as a function of waist position, for two different waist sizes. The optical positions of the four interaction zones are marked by solid vertical lines. (b) The Fisher information, $F$, as a function of waist position, for two different waist sizes. The average interaction position is marked by a solid vertical line. }
\end{figure}

In the limit $N \to \infty$ the Fisher information per atom as a function of detuning is $F=(dP/d\Delta)^2/P$ (see Appendix \ref{app:F}), which is maximised on the slopes of the central fringe. 
Assuming sinusoidal fringes with no phase shift sitting on a flat background, $db/d\Delta=0$, the maximum gradient around resonance, $\Delta=\delta$, is $dP/d\Delta \approx T_f b_0c_0$, where $1/T_f$ is the period of the fringes. This gives
\begin{equation}
F\approx T_f^2 \frac{b_0c_0^2}{1+c_0}
\end{equation}
as the Fisher information on resonance.
For equal pulse areas at each atom-laser interaction and fixed velocities, $v$ and $v_z$, the Ramsey background and fringe envelopes are identical $a_+=a_-=a(v,v_z)$, such that the RB background contribution from each recoil component is $(a-a^2)$ and each fringe envelope is $a^2/2$. 
The brightness and contrast then become $b_0=2\langle a \rangle (1-r)$ and $c_0=r/4(1-r)$, where $r=\langle a^2 \rangle/\langle a \rangle$ and $a$ is evaluated at resonance.
For $\pi/2$ pulses $a$ is a flat top peak $\sim \exp(- v_z^4)$ as a function of $v_z$, with height $1/2$. We show this in Appendix \ref{app:RI}. For fixed $v$ and narrow Doppler distribution $\rho(v_z) \sim \delta(0)$ we recover the theoretical maximum contrast of the central fringe, $c_0=1/4$, since $r=a=1/2$. 
This gives Fisher information $F_0=T^2_f/40$, which we use as our unit for the general Fisher information $F$.
In the opposite limit of a flat Doppler distribution $\rho(v_z) \sim \mathrm{const.}$, and the quartic decay of $a$ gives $r=1/2^{5/4}$, resulting in a Doppler limited maximum contrast $c\approx 0.18 $.
As noted in \cite{PhysRevA.30.1836}, the Doppler limited contrast is partially improved by using larger target pulse areas $A>\pi/2$.

 In Fig.~\ref{fig:4}(a) we plot the brightness, $b_0$, and lower recoil contrast, $c_0$, at resonance as a function of the position of the lasers focal point $l_w$ on the optical path. 
 We use waist radii $w_0= \qtty{0.125}{mm}$ and $w_0= \qtty{0.2}{mm}$, and target pulse area $A=\pi/2$ for atoms travelling at mean velocity through the waist.
 Vertical black lines indicate when the focal point aligns with an interaction zone $l_w=l_i$. 
 Atoms intersect the $i$th interaction zone in ascending order, and from left to right in Fig.~\ref{fig:4}(a) the vertical lines indicate the positions along the optical axis of the $1$st, $4$th, $2$nd, and $3$rd atom-laser interactions, $l_1$, $l_4$, $l_2$, and $l_3$, respectively, whose values are defined in the previous section.
 These are not in ascending order due to the optical path being folded in a spiral (see Fig.~\ref{fig:1}) so the atomic trajectory is not in one-to-one correspondence with the optical trajectory.
 For the smaller waist, the brightness has peaks with width $z_R=\qtty{7.5}{cm}$ as the focal point coincides with the four interaction zones $l_w=l_i$. 
 For the larger waist the peaks have width $z_R=\qtty{19}{cm}$,  similar to the optical path lengths that separate the interactions, so the peaks merge. 
 Aligning the focal point with an interaction zone maximises the probability that it causes the transition, $\ket{g_0} \to \ket{e_{\pm 1}}$. 
 Brightness is maximised with laser focus on the final interaction, $l_w=l_4=\qtty{0.3}{cm}$, because it enhances the contribution from lower recoil atoms that have otherwise remained in the ground state $\ket{g_0}$, suffering no loss due to atomic decay. 
 Conversely, aligning with the first interaction, $l_w=l_1=\qtty{0}{cm}$, enhances the atomic paths that suffer most decay, making it the smallest peak.
 
 Since $c_0\propto 1/b_0$, the contrast in Fig~\ref{fig:4}(a) has minima of width $\sim z_R$ when $b_0$ is maximised, i.e. when the laser is focused at one of the four interaction zones, and maxima in between interactions where the pulse areas become more homogeneous. 
 For the larger waist size these features smooth out and the magnitude is approximately double that of the smaller waist, since the lower radius of curvature of the wavefronts leads to increased pulse area homogeneity. 
 Despite the complex behaviour of $b_0$ and $c_0$, the Fisher information plotted in Fig~\ref{fig:4}(b) for small and large waists each exhibit a unique maximum. 
 For both waist sizes $F$ is maximised near the geometric mean optical position of the interactions, $\overline{l}=\sum_il_i/4=\qtty{0.395}{cm}$, which is marked as a vertical line. 

\begin{figure}
\begin{centering}
\includegraphics[scale=1]{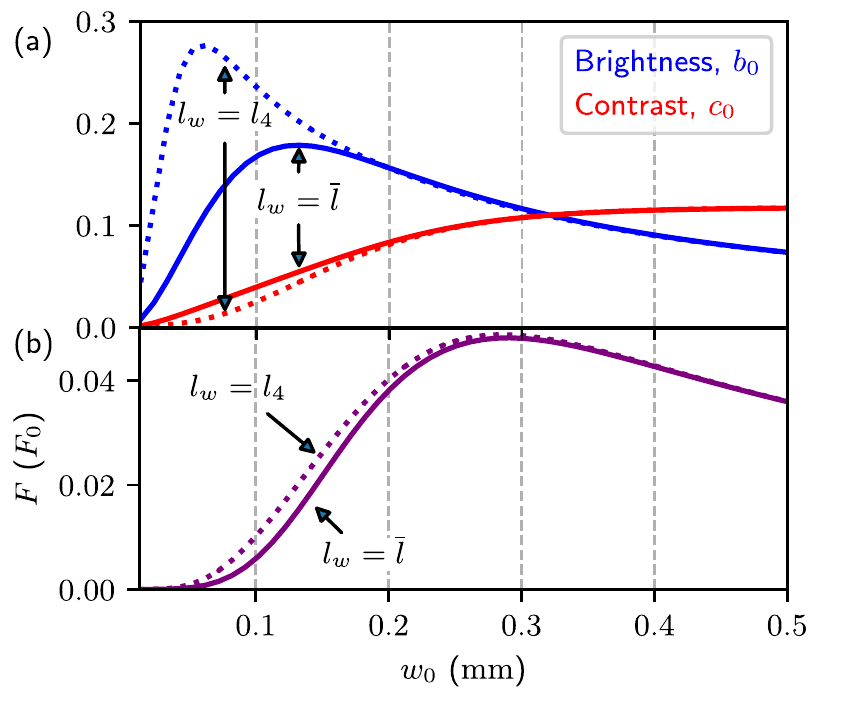}
\par\end{centering}
\caption{\label{fig:5} (a) The brightness, $b_0$, and contrast, $c_0$, at the central lower recoil fringe and (b) The Fisher information, $F$, as a function of waist position, for two different waist sizes.   }
\end{figure}
 
 In Fig.~\ref{fig:5}(a) we plot the brightness and contrast as a function of waist radius $w$ using waist positions that maximise the Fisher information ($l_w=\overline{l}$) and the brightness ($l_w=l_4$). 
 As the waist size tends to zero the wavefronts become too curved for the atoms to be resonant with any significant portion of the laser so the brightness and contrast, and hence the whole signal, vanishes.
 For larger waists, the wavefronts flatten and pulse areas homogenise, leading to an improved contrast. 
 The contrast does not improve to the Doppler limit $c_0 \approx 0.18$ since atomic decay and the spread of longitudinal velocity cause additional degradation. 
 In the large waist limit, longer atomic transit times reduce the portion of the $v_z$ distribution that contribute to the signal, reducing brightness. 
 Competition between wavefront curvature and time-of-flight effects therefore leads to a unique maximum in the brightness. 
 The Fisher information shown in Fig.~\ref{fig:5}(b) reaches a maximum value $F/F_0\approx 0.05$ at $w_0 \approx \qtty{0.3}{mm}$ for both waist positions used. 
 This optimised value of $w_0$ is similar to that used in \cite{PhysRevLett.123.073202}, and corresponds to a Rayleigh range $z_R \approx \qtty{43}{cm}$, similar to the largest optical path length in the system, $l_4-l_1 =\qtty{42}{cm}$.

The interplay of wavefront curvature and Doppler effects that determines the maxima of $F$ in Fig~\ref{fig:5}(b) suggests a matching condition between the angular divergences of the laser, $\alpha_l=w_0/z_R$, and atomic beam, $\alpha_a=v_w/v$ where $v_w$ is the width of the transverse velocity distribution.
This effect is hard to elucidate using the full RB model, so to understand this further we consider atoms with fixed longitudinal velocity $v$ passing through a single laser with pulse area, $2\Phi$, given by Eq.~\eqref{eq:MEPA} at a fixed distance from the waist, $z$.
For low pulse area, the excitation probability is proportional to the pulse are squared, $p \propto \Phi^2$.
Since $\Phi$ is Gaussian in $v_z$ we can find the average over $v_z$, $\langle \Phi^2 \rangle$, analytically. 
Using $\rho(v_z)\propto \exp(-v_z^2/v_w^2)$, we find the excitation probability is maximised when $\alpha_l/\alpha_a=\sqrt{(z_R/z)^2-1}$.
Since the divergences positive numbers this says that, for fixed $z$, the optimised laser parameters must strictly have $z_R\geq z$. 
In the limit of a flat Doppler distribution, $\alpha \propto v_w \to \infty$, this condition simplifies to $z_R=z$. 
This implies the excitation probability of a single atom-laser interaction is maximised when the waist size is chosen such that atoms intersect the laser at the point of maximum wavefront curvature.

\section{Frequency Stability}
\label{sec:Freq}

Fluctuations in the frequency offset of the measured central fringe around resonance lead to instability in the measured clock transition frequency.
Using Eq.~\eqref{eq:MEPh}, for an atom with fixed velocity the lower recoil fringes around resonance have total phase
\begin{equation}\label{eq:Tph}
  \theta_L = 2T\left(1 +\frac{l_s}{d_r} \frac{v_z}{v}\right)(\Delta-\delta) + kl_s\left(1-\frac{v_z^2}{2v^2}\right) +\frac{g_s}{2},
\end{equation}
where $l_s=(l_1-l_2)+(l_3-l_4)$ is the sum of optical path lengths between the interactions in each RI. The total Gouy phase contribution is given by $g_s$ and is the only term that depends on either the position or size of the waist. 
We have neglected terms of order $(\Delta/kv)^2$ which are vanishing near the central fringe and from here on assume the atomic beam is perpendicular to the lasers such that $\langle v_z \rangle =0$.

Upon inspection of the spatial phase contribution to Eq.~\eqref{eq:Tph}, which is $\propto kl_s$, velocity averaging leads to dephasing of the fringes if the phase shift is $\langle v_z^2 /v^2 \rangle k l_s/2>\pi$. 
The geometry of our model system leads to typical values $\langle v_z^2/v^2 \rangle \sim v_w^2/v_m^2 \sim 10^{-6}$ and our path length sum is $l_s = \qtty{2}{cm}$, amounting to a phase shift of only $\sim \pi/60$.
For fringes with period $\sim \qtty{3}{kHz}$ this translates to frequency shifts $\sim \qtty{25}{Hz}$, equivalent to a fractional frequency shift from the the true clock transition $\sim \num{e-13}$.

The first term in Eq.~\eqref{eq:Tph} describes a correction, $v_z l_s/v d_r \sim \num{e-4} $, to the Ramsey time between interactions, $T$.
This is because the atoms are not excited exactly at the optical axes at each interaction, highlighted in Fig.~\ref{fig:1}, but are instead excited slightly earlier or later due to the resonant wavevector being off-axis, leading to differing times of flight between interactions.
To leading order in $v_z/v$, for our system these time of flight corrections result in corrections to the period also of order $\num{e-4}$, corresponding to a sub-Hertz frequency shift of the central fringe. 

The Gouy phase term, $g_s$, in Eq.~\eqref{eq:Tph}, is a sum of four terms $\pm \arctan((l_w-l_i)/z_R)$ and so can in principle cause full period shifts $g_s/2\leq 2 \pi$. 
For our system, with small Rayleigh ranges $z_R \sim \qtty{10}{cm}$, the maximum Gouy phase shift is $\sim \pi/4$ which corresponds to a frequency shift $\sim \qtty{400}{Hz}$. 

 \begin{figure}
\begin{centering}
\includegraphics[scale=1]{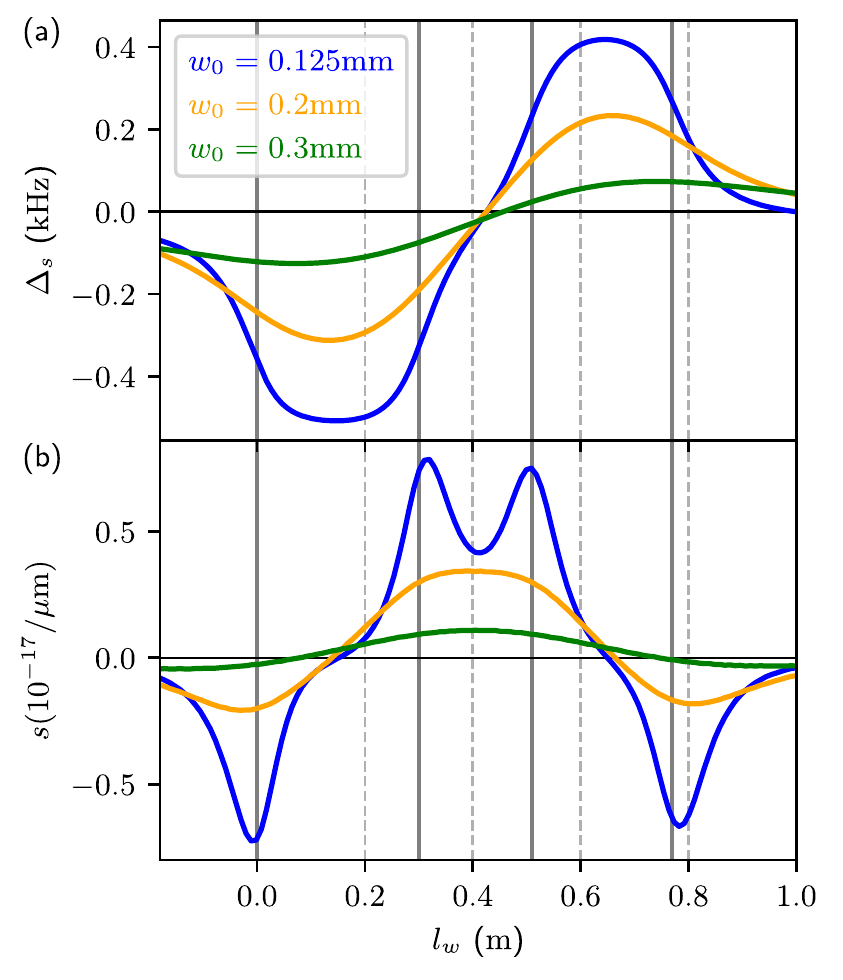}
\par\end{centering}
\caption{\label{fig:6} (a) Frequency shift, $\Delta_s$, of the lower recoil fringes and (b) fractional stability, $s$, of the shift with respect to fluctuations in the waist position, $l_w$, as a function of waist position, $l_w$, for three different waist sizes $w_0$. 
The positions, $l_i$, of the interactions along the optical path, are marked by solid vertical lines.}
\end{figure}

To find the total fringe frequency shift in the velocity averaged signal, Eq.~\eqref{eq:P}, we calculate the fringe envelope function and factor it out to leave undamped fringes with amplitude $1$. 
These fringes generally have frequency modulations, but near resonance we can fit them to $\cos(2T_f(\Delta-\Delta_s))$ to obtain the effective frequency shift, $\Delta_s$ and period, $1/T_f$.
In Fig.~\ref{fig:6}(a) we plot $\Delta_s$ as a function of waist position, $l_w$, for various waist sizes.
The positions of the interaction zones are marked with vertical lines.
The Gouy shift is the dominant contribution to $\Delta_s$, as has been found in other types of atom interferometry \cite{Morel2020}, with each contribution $\pm\arctan((l_i-l_w)/z_R)$ having slope centered at the corresponding interaction, $l_w=l_i$, with gradient $\sim \pm1/z_R$. 
For the smallest waist the largest shifts are $\sim \qtty{400}{Hz}$.
For the largest waists and larger $z_R$, the gradients get smaller leading to an overall reduction in the size of the shift. 
For the largest waist, with Rayleigh range $z_R \sim \qtty{40}{cm}$, the shift is still $\sim \qtty{100}{Hz}$. 
As well as Gouy contribution to $\Delta_s$, there is also an approximately constant frequency shift $\sim-\qtty{50}{Hz}$ due to the spatial phase term $\propto kl_s$ in Eq.~\eqref{eq:Tph}.

The waist position in Fig.~\ref{fig:6}(a) that gives net zero shift, $\Delta_s=0$, is $l_w \approx \qtty{40}{cm}$ similar to the mean optical position of the interactions, $\overline{l}\approx \qtty{40}{cm}$. 
This is similar to the waist position that maximises contrast and Fisher information shown in Fig.~\ref{fig:4}. 
In theory this is therefore a good configuration to minimise error in the meausred clock transition.
However, it may be more desirable to choose the waist position such that the sensitivity to waist position is minimized, since fluctuations in the optical path lengths will lead to fluctuations in the distances from the waist at each interaction zone and thus fluctuations in the measured frequency.
Sensitivity of the frequency shifts to fluctuations in $l_w$ can be estimated from the derivative of $\Delta_s$. Thus we define the fractional frequency stability with respect to the position of the waist as
\begin{equation}
s= \frac{1}{\omega_c}\frac{d\Delta_s}{dl_w}
\end{equation}
where $\omega_c$ is the bare clock transition frequency. 
We plot $s$ in Fig.~\ref{fig:6}(b), showing the typical fractional frequency shift in clock frequency due to $\sim 1\, \mu \mathrm{m}$ fluctuations in the position of the waist, $l_w$.

Since the Gouy shift is a sum of four arctangents, the sensitivity, $s$, is a sum of four Lorentzians centered at the interactions, $l_w=l_i$, with width $\sim z_R$ and height $\propto 1/z_R$. 
For the smallest waist, with $z_r=\qtty{7}{cm}$, the sensitivity is largest as the interaction zones, reaching fractional instabilities of $\sim 10^{-17}$ per $\mu$m fluctuation in $l_w$. 
For the two larger waists, with Rayleigh ranges $z_R = \qtty{19}{cm}$ and $z_R=\qtty{43}{cm}$, the peaks begin to overlap. 
The largest waist size we use is $w_0=\qtty{0.3}{mm}$, which optimises the Fisher information with respect to $w_0$, as shown in Fig.~\ref{fig:5}(b), and gives a sensitivity $s \sim 10^{-18}$ when the focal point is optimised to give best contrast, $l_w\approx\qtty{40}{cm}$. 
Instability is suppressed in Fig.~\ref{fig:6}(b) when the waist is positioned near the midpoint of the optical paths between either the first and last interactions, $l_w=(l_1+l_4)/2\approx \qtty{15}{cm}$, or the second and third, $l_w=(l_2+l_3)/2\approx \qtty{67}{cm}$.

A method to partially eliminate residual frequency shifts, employed in \cite{PhysRevLett.123.073202}, is to use counterpropagating atomic beams, where each beam traverses the same interaction zones but in reverse order. 
For an atom with spatial trajectory defined by $v$ and $v_z$, the fringe phase for the corresponding counterpropagating atom is found by setting $l_s \to -l_s$ and $g_s \to - g_s$ in Eq.~\eqref{eq:Tph}. 
In powers of $v_z/v$, the lowest order correction to the frequency shift from our theory that is not cancelled by use of this method is $ g_s/T \left(v_z l_s/4vd_r \right)^2 \sim 10 \, \mu \mathrm{Hz}$. This implies that, for the system we consider, the counterpropagating beam method can at best achieve fractional frequency instabilities of $\sim \num{e-20}$.

\section{Conclusion}
Motivated by recent experiments implementing compact optical Ramsey-Bord\'{e} interferometry \cite{PhysRevLett.123.073202}, we have developed an analytic model for optical atomic beam clocks that accounts for laser wavefront curvature. 
The results reported in \cite{PhysRevLett.123.073202} indicate significant sensitivity of the measured clock signal to the geometry and collimation of the lasers. 
Our model confirms this, showing that varying levels of wavefront curvature at the interaction zones results in pulse area inhomogeneity and a complicated variation of fringe contrast with respect to the position of the focal point. 
As such, correct positioning of the waist of the laser is crucial to optimising the clock signal. 

By characterising the signal by its Fisher information, we rigorously identify an unambiguous optimal waist position and size which ensure the best possible signal quality. The optimal waist positions turns out to be approximately the point of highest symmetry; the average position of the atom-laser interactions. This configuration minimises inhomoegneity of the pulse areas at each interaction leading to higher contrast fringes than if the laser was focused at one of the interaction zones.

Simply increasing the waist size of the laser, thereby eliminating wavefront curvature and pulse area inhomogeneity, does not necessarily improve the signal. This is because it also increases the atom-laser time of flight and reduces the proportion of Doppler shifted atoms in the thermal beam that can contribute to the signal.
Having a smaller waist, and hence some degree of wavefront curvature, does increase the proportion of Doppler shifted atoms that contribute to the signal but this is purely due to the smaller transit time and not because a spread of wavevectors addresses more atoms.
In the limit of an extremely small waist the fraction of the highly curved wavefront that the atom is resonant with is vanishing, so in this limit the signal vanishes.
The optimal waist size we identify is relatively small, giving an optimal laser configuration that is uncollimated.
This is consistent with the optimal laser configuration reported in \cite{PhysRevLett.123.073202}.

We have identified a number of sources of fringe frequency shifts that arise from using a realistic Gaussian laser.
Most prominent of these is the Guoy phase, which varies at each interaction zone when using a folded laser geometry and can lead to fractional frequency of order $\num{e-12}$.
We also found smaller contributions to the frequency shift arising from time of flight corrections and spatial phase variations, both of which are a direct consequence of the curved wavefronts in the Gaussian laser.
Interestingly, all such frequency shifts, including the Guoy phase, are minimised by positioning the waist of the laser at a similar point to where the Fisher information is optimised.
We have shown, however, that this configuration also makes the shifts unstable to perturbations in the position of the waist, with $\sim 1\mu \mathrm{m}$ fluctuations in the waist position giving a fractional frequency instability of order $\num{e-17}$. The waist can alternatively be positioned such that this sensitivity vanished, but at the expense of a sub-optimal Fisher information and the introduction of frequency shifts of order $\qtty{0.5}{kHz}$.

As optical atomic clocks are made more compact and portable, constraints on the system complexity must be employed and their consequences well understood.
Our results provide analytic insight into how the effects of wavefront curvature manifest in compact optical thermal beam clocks, and provide an efficient way of modeling the observable velocity averaged signal. We have shown how the laser geometry in these systems plays a crucial role in optimising the signal. Our analyses and the optimisation methods we have used will aid in the design of next generation compact optical atomic clocks.

\section{Acknowledgements}

 This research was supported by the Australian Research Council Centres of Excellence for Engineered Quantum Systems Projects No. CE170100009. This research was funded through a Quantum Technologies Research Network Grant through the Defence Science \& Technology Group, Australian Government (QT71).
 
\appendix 

\section{Ramsey Background}
\label{app:RI}

\begin{figure}
\begin{centering}
\includegraphics[scale=1]{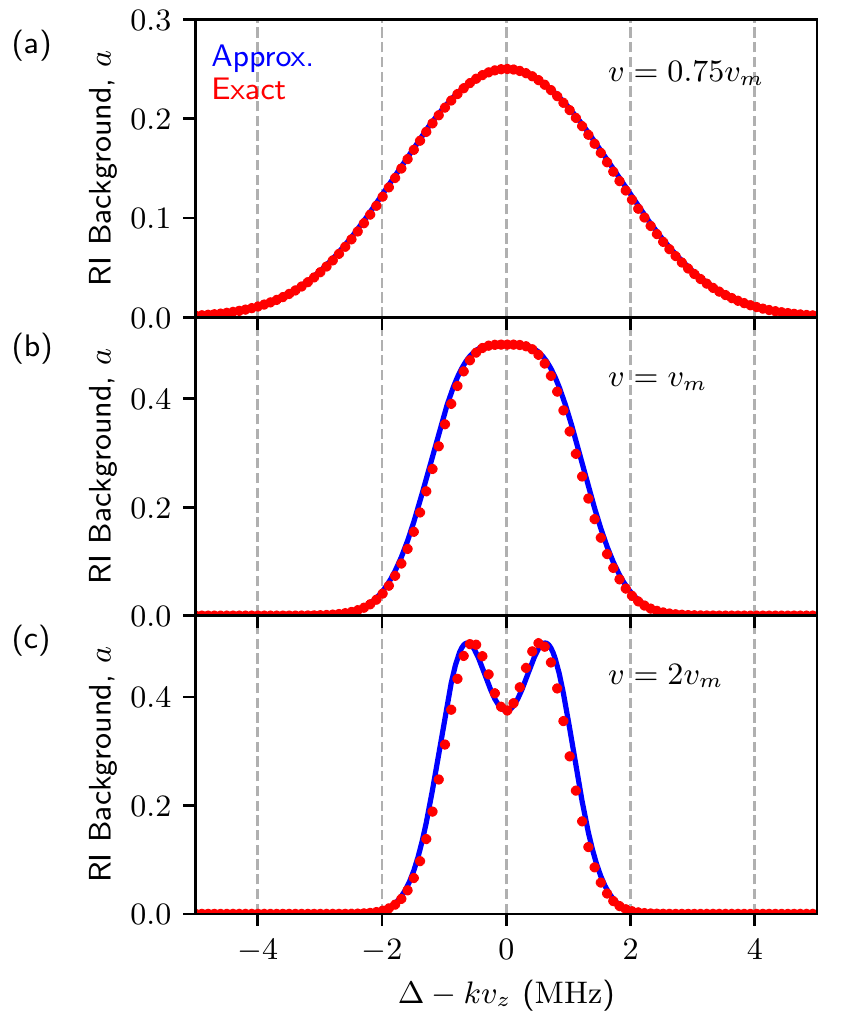}
\par\end{centering}
\caption{\label{fig:a} The background function for a Ramsey interferometer for atoms with fixed transverse velocity $v_z$, for various longitudinal velocity $v$. We compare the approximate first order Magnus expansion with exact numerics, finding perfect agreement for slow atoms and good agreement for faster atoms. }
\end{figure}

For a single Ramsey interferometer (RI) with identical pulse areas at each interaction zone the background, $a_+$, and fringe envelope, $a_-$, given by Eq.~\eqref{eq:apm} become identical, $a_+=a_- =a$. We plot this function in Fig.~\ref{fig:a}, using both the approximate solution Eqs.~\eqref{eq:MEPA} and \eqref{eq:MEPh} for the pulse area and the exact numerics. We assume fixed transverse velocity $v_z$ and vary the longitudinal velocity $v$. We have assumed the laser waist is imaged exactly to each of the two atom-laser interaction zones, meaning the laser is effectively a plane wave. The laser intensity is normalised such that on resonance the atom receives the target pulse area $2\Phi = A= \pi/2$ when $v=v_m$, seen in Fig.~\ref{fig:a}(b). In this case the background has a flat top form $a \sim \exp(-(\Delta-kv_z)^4)$, which is accurately reproduced by the approximate solution.

Atoms with larger longitudinal velocity, $v>v_m$, receive pulse areas $2\Phi<\pi/2$, leading to a small background. This is seen in Fig.~\ref{fig:a}(a), where the pulse area is $2\Phi=\pi/4$ on resonance and the background has a Gaussian profile. The background is also broader than in Fig.~\ref{fig:a}(b) because of the shorter atom-laser transit time. In this regime the approximate and exact solutions are essentially identical. Atoms with smaller longitudinal velocity, $v<v_m$, receive pulse areas $2\Phi>\pi/2$ on resonance, leading to a Lamb dip appearing in the background. This is seen in Fig.~\ref{fig:a}(c). The background is narrower than in Fig.~\ref{fig:a}(b) because of the longer atom-laser transit time. This is the strong driving regime where the approximate solution begins to breakdown, but still captures the essential features of the exact background. \newline
\vfill \eject

\section{Fisher Information}
\label{app:F}
The Fisher information is defined as
\begin{equation}
    I(\Delta)=\int \left(\frac{d}{d\Delta} \log f(x|\Delta) \right )^2f(x|\Delta) dx,
\end{equation}
where $f(x|\Delta)$ is the probability of obtaining the outcome $x$ given $\Delta$. Here $x$ is the measured number of excited state atoms and $\Delta$ is the laser detuning.
We assume $x$ is Poisson distributed with mean $n=n(\Delta)$, such that
\begin{equation}
    f(x|\Delta) = \frac{n^x}{x!} e^{-n}.
\end{equation}
Since $n \sim O(\num{e12})$ \cite{PhysRevLett.123.073202} this can be approximated as a Gaussian
\begin{equation}
   f(x|\Delta) \approx  \frac{1}{\sqrt{2\pi n}} e^{-\frac{(x-n)^2}{2n}}.
\end{equation}
We move to the continuum limit by writing $x=pN$ and $n=PN$, where $N$ is the total number of atoms, $p=p(\Delta)$ is the fluctuating excited state population of the $N$ atoms, and $P=P(\Delta)$ is the mean excited state population. 
Changing variables from $x$ to $p$, the Gaussian becomes
\begin{equation}
   f(p|\Delta) \approx  \frac{1}{\sqrt{2\pi PN}} e^{-\frac{N(p-P)^2}{2P}}
\end{equation}
which has width $\propto 1/\sqrt{N}$. 
Using this in the formula for the Fisher information we get
\begin{equation}
    I(\Delta)=\frac{1}{2}(1-2 N P(\Delta)) \left(\frac{P'(\Delta)}{P(\Delta)} \right)^2.
\end{equation}
In the limit $N \to \infty$ the Fisher information per atom is thus
\begin{equation}
    F(\Delta)=\lim_{N \to \infty} \frac{I(\Delta)}{N} =\frac{P'(\Delta)^2}{P(\Delta)}.
\end{equation}
At the peak of the central lower recoil fringe we have $P'(\Delta=\delta)=0$ so $F(\Delta)$ is instead maximised at the point of steepest slope either side of the fringe.
Given that the amplitude of the central fringe is the brightness multiplied by the contrast, $b_0c_0$, the gradient of the slope can be approximated as the amplitude divided by the fringe width which we define as $1/T_f$. This gives
\begin{equation}
    P' \approx T_f b_0c_0.
\end{equation}
around the central fringe.
The mean excited state population at the central fringe is
\begin{equation}
    P = b_0(1+c_0).
\end{equation}
The Fisher information per atom contained in the central fringe can then be estimated as
\begin{equation}
   F \approx T_f^2 \frac{b_0c_0^2}{1+c_0}
\end{equation}

\bibliographystyle{apsrev4-1} 
\bibliography{bibliography.bib} 

\end{document}